\documentclass[11pt]{article}

\usepackage{latexsym}
\usepackage{amsfonts}

\newtheorem{theorem}{Theorem}

\newtheorem{corollary}[theorem]{Corollary}

\newenvironment{proof}{\trivlist\item[\hskip \labelsep{\textit{Proof.}}]}%
                 {\hfil\hfill$\square$\endtrivlist}

\begin{document}

\title{Algebraic and differential Rainich conditions for symmetric trace-free tensors
of higher rank}

\author{G Bergqvist$^1$ and P Lankinen$^2$ \\
${}^1$Matematiska institutionen, Link\"opings universitet, \\
      SE-581 83 Link\"oping, Sweden \\
${}^2$Department of Mathematics and Physics, M\"alardalen University, \\
SE-721 23 V\"aster{\aa}s, Sweden \\
gober@mai.liu.se, paul.lankinen@mdh.se}

\maketitle

\begin{abstract}
We study Rainich-like conditions for symmetric and trace-free tensors T. For arbitrary
even rank we find a necessary and sufficient differential condition for a tensor
to satisfy the source free field equation. For rank 4, in a generic case, we combine these
conditions with previously obtained algebraic conditions to 
 obtain a complete set of algebraic and differential conditions on T for it to be a superenergy
  tensor of a Weyl candidate tensor satisfying the Bianchi vacuum equations. 
By a result of Bell  and Szekeres this implies that in vacuum, generically, 
T must be the Bel-Robinson tensor
  of the spacetime. For the rank 3 case we derive a complete set of necessary
 algebraic and differential conditions for T to be the superenergy tensor of a massless spin 3/2
 field satisfying the source free field equation.
 \end{abstract}




\newpage

\section{Introduction}

Given a symmetric trace-free divergence-free tensor $T_{ab}$ satisfying the
dominant energy condition ($T_{ab}u^a v^b\ge 0$ for all future-directed
causal vectors $u^a$ and $v^a$), one can ask what more is required of
$T_{ab}$ for it to be the energy-momentum tensor of some given
physical field. It turns out that to completely characterize
$T_{ab}$ we will need both an algebraic and a differential
condition. Assuming dimension 4 and Lorentzian metric,  the
following is a result in classical Rainich-Misner-Wheeler theory
\cite{MW,PR,R}:

\smallskip

\begin{theorem}\label{th:RaiEM}
A symmetric trace-free tensor $T_{ab}$ which satisfies the dominant energy condition
can be written $T_{ab}=-{1\over 2}(F_{ac}F_{b}{}^c+{}^*F_{ac}{}^*F_{b}{}^c)\equiv -F_{ac}
F_{b}{}^c+{1\over 4} g_{ab}F_{cd}F^{cd}$, where $F_{ab}$
is a 2-form, if and only if
\begin{equation}    \label{RaiEM}
T_{ac}T_{b}{}^c={1\over 4}g_{ab}T_{cd}T^{cd}\ .
\end{equation}
\end{theorem}

\smallskip
Here ${}^*F_{ab}$ is the dual 2-form of $F_{ab}$. Removing the assumption of
the dominant energy condition Theorem \ref{th:RaiEM} is still true up to sign  \cite{BS} :
$\pm T_{ab}=-{1\over 2}(F_{ac}F_{b}{}^c+{}^*F_{ac}{}^*F_{b}{}^c)$ if and only if
(\ref{RaiEM}) is satisfied.

A tensor $T_{ab}$ fulfilling the requirements of the theorem is
algebraically the energy-momentum tensor of a Maxwell field
$F_{ab}$. Equivalently a tensor satisfying the given requirements
can be written $T_{ab}=2\varphi_{AB}\bar{\varphi}_{A'B'}$ where
$\varphi_{AB}$ is a spinor representing the Maxwell field.
In the theorem $F_{ab}$ is only determined up to a duality rotation
$F_{ab}\to F_{ab}\cos\theta + {}^*F_{ab}\sin\theta$ which corresponds to
$ \varphi_{AB}\to e^{-i\theta}\varphi_{AB}$.

Of course we will have to accompany this algebraic condition with
a differential condition that assures that the field $F_{ab}$ (or
equally $\varphi_{AB}$) satisfies the source-free Maxwell's equations. The following
is known \cite{MW,PR,R}

\begin{theorem}     \label{ClassDiffRai}
Suppose that $T_{ab}=-{1\over 2}(\tilde F_{ac}\tilde F_{b}{}^c+
{}^*\tilde F_{ac}{}^*\tilde F_{b}{}^c)$ for some
2-form $\tilde F_{ab}$ and that $\nabla^a T_{ab}=0\neq  T_{ab}T^{ab}$. 
Then  $T_{ab}=-{1\over 2}(F_{ac}F_{b}{}^c+{}^*F_{ac}{}^*F_{b}{}^c)$ for 
some 2-form $F_{ab}$ satisfying the source-free Maxwell equations
$\nabla_{[a}F_{bc]}=0=\nabla_a F^{ab}$, if and only if
\begin{equation}    \label{Sint}
\nabla_b S_a=\nabla_a S_b \qquad {\mbox{where}}\qquad
S_{c}=\frac{e_{ac}{}^{pq}T_{bp}\nabla_qT^{ab}}{T_{ef}T^{ef}}
\end{equation}
\end{theorem}
Note that $F_{ab}$ is obtained from $\tilde F_{ab}$ by a duality rotation and
that the source-free Maxwell equations in spinor form are
just $\nabla^{AA'}\varphi_{AB}=0$ \cite{PR}. Using spinors, Theorem
\ref{ClassDiffRai} can equivalently be written
\begin{theorem}     \label{ClassDiffRaiSp}
Suppose that $T_{ab}=2\phi_{AB}\bar{\phi}_{A'B'}$ for some
symmetric spinor $\phi_{AB}$ and that $\nabla^a T_{ab}=0\neq  T_{ab}T^{ab}$. Then
$T_{ab}=2\varphi_{AB}\bar{\varphi}_{A'B'}$ for some
symmetric spinor $\varphi_{AB}$
satisfying $\nabla^{AA'}\varphi_{AB}=0$ if and only if (\ref{Sint}) is satisfied.
\end{theorem}
The validity of Theorems  \ref{ClassDiffRai} and  \ref{ClassDiffRaiSp} is
obviously restricted to cases where
$T_{ab}T^{ab}\neq 0$, i.e. when the two principal null directions
of $\varphi_{AB}$ are different (non-null electromagnetic fields).
In the null case results cannot be stated in an equally simple way, see \cite{G,Lu}.

Theorems \ref{th:RaiEM} and \ref{ClassDiffRai} imply
\begin{corollary}     \label{ClassRai}
A symmetric trace-free and divergence-free tensor  $T_{ab}$ with $T_{ab}T^{ab}\neq 0$ 
is, up to sign, the energy-momentum tensor of a source-free Maxwell field if and only if
 (\ref{RaiEM}) and  (\ref{Sint}) are satisfied.
\end{corollary}
If one uses Einstein's equation, $T_{ab}$ may be replaced by the Ricci tensor $R_{ab}$
in the corollary since $T_{ab}$ is trace-free. In this case the Ricci tensor is automatically
divergence-free so this is not needed as a condition. Equations  (\ref{RaiEM}) and (\ref{Sint}) 
are then satisfied for $R_{ab}$ if and only if $R_{ab}$ is the Ricci tensor for an 
Einstein-Maxwell spacetime.

The algebraic result of Theorem \ref{th:RaiEM} has been generalized
 to arbitrary dimension
and arbitrary trace of $T_{ab}$ when (\ref{RaiEM})  is assumed \cite{BS},
and to cases in higher dimension
when  (\ref{RaiEM}) is replaced by a third-order equation for $T_{ab}$ \cite{BH}.
In these generalizations only rank-2 tensors $T_{ab}$ were considered.
We will here generalize Theorems \ref{th:RaiEM} and \ref{ClassDiffRai} to include symmetric
trace-free tensors of rank 3 and 4. For higher rank tensors the dominant energy
condition is replaced by a generalization called the {\it{dominant
property}},

\begin{equation}    \label{dp}
    T_{a_1...a_r}u_1^{a_1}...u_r^{a_r}\geq 0
\end{equation}
for all causal vectors $u_k^{a_k}$. The spacetime dimension will always
be four here and the metric will be assumed to be of Lorentzian signature.
The methods will be spinorial
so that we will start by reviewing necessary facts about these.
After that a differential condition for symmetric trace-free and divergence-free
tensors of even rank is obtained, generalizing Theorem \ref{ClassDiffRai},
and applied to the Bel-Robinson
tensor. The algebraic condition for the Bel-Robinson tensor was already obtained
in \cite{BL} and we can now give a complete characterization of the
Bel-Robinson tensor. The Bel-Robinson tensor is the so-called superenergy
tensor of the Weyl tensor or  the Weyl spinor. To any tensor on
a Lorentzian manifold there is a corresponding superenergy tensor of even rank
and this always has the dominant property  \cite{B,S}. In \cite{LP} this definition
was extended to include also superenergy tensors of spinors, which may then
be of odd rank. Here we derive both algebraic and
differential conditions on symmetric trace-free and divergence-free tensors of rank 3,
giving a complete characterization of superenergy tensors  of
massless spin-$\frac{3}{2}$ fields. 

\section{Some useful spinor identities}\label{sec:spinors}

We review some well-known facts about spinors that will be
important to us. The formulas can be found in the book by Penrose
and Rindler \cite{PR} and we also follow their notation and
conventions (except for a factor 4 in the definition of the
Bel-Robinson tensor). Spinor expressions for general superenergy
tensors are given in \cite{B}.

We use capital letters $A,B,\dots ,A',B',\dots$ for spinor indices
and identify with tensor indices $a,b,\dots$ according to $AA'=a$.
A spinor $P_{AB\mathcal{Q}}$ , where $\mathcal{Q}$ represents some
set of spinor indices, can be divided up into its symmetric and
antisymmetric parts with respect to a pair of indices
$$
P_{AB\mathcal{Q}}={1\over 2}(P_{AB\mathcal{Q}}+P_{BA\mathcal{Q}})+
{1\over 2}(P_{AB\mathcal{Q}}-P_{BA\mathcal{Q}})=P_{(AB)\mathcal{Q}}+P_{[AB]\mathcal{Q}} \ .
$$
The antisymmetric part can be written
$$
P_{[AB]\mathcal{Q}}={1\over 2}\varepsilon_{AB}P_{C}{}^C{}_{\mathcal{Q}} \ ,
$$
where $\varepsilon_{AB}=-\varepsilon_{BA}$, so
\begin{equation}\label{eq:sas}
P_{AB\mathcal{Q}}=P_{(AB)\mathcal{Q}}+{1\over 2}\varepsilon_{AB}P_{C}{}^C{}_{\mathcal{Q}}\ .
\end{equation}
From this one also has
\begin{equation}\label{eq:sas2}
P_{AB\mathcal{Q}}=P_{BA\mathcal{Q}}+\varepsilon_{AB}P_{C}{}^C{}_{\mathcal{Q}} \ .
\end{equation}
A simple but very useful rule is
\begin{equation}\label{eq:zz}
P_{C}{}^C{}_{\mathcal{Q}}=-P^C{}_{C\mathcal{Q}} \ .
\end{equation}
Note that if $P_{ab\mathcal{Q}}=P_{ba\mathcal{Q}}$ then we have
$$
P_{BAA'B'\mathcal{Q}}=P_{ab\mathcal{Q}}-{1\over 2}g_{ab}P_{c}{}^{c}{}_{\mathcal{Q}} \ ,
$$
where $g_{ab}=\varepsilon_{AB}\bar\varepsilon_{A'B'}$ ; so permuting $A$ and $B$ gives a trace reversal. From this we find another formula
we shall need (with $P_{ab\mathcal{Q}}$ not necessarily symmetric in $ab$)
\begin{equation}\label{eq:ds}
P_{(AB)(A'B')\mathcal{Q}}=P_{(ab)\mathcal{Q}}-{1\over 4}g_{ab}P_{c}{}^{c}{}_{\mathcal{Q}} \ .
\end{equation}
The completely antisymmetric tensor $e_{abcd}$, normalized by
$e_{abcd}e^{abcd}=-24$, can be written

$$
    e_{abcd}=i\varepsilon_{AC}\varepsilon_{BD}\varepsilon_{A'D'}\varepsilon_{B'C'}-i\varepsilon_{AD}\varepsilon_{BC}\varepsilon_{A'C'}\varepsilon_{B'D'}
$$
Raising the indices $cd$ and
applying this tensor to the tensor $P_{cd\mathcal{Q}}=P_{CC'DD'\mathcal{Q}}$ gives the
following useful relation
\begin{equation}    \label{changepairofindices}
    e_{AA'BB'}{}^{CC'DD'}P_{CC'DD'\mathcal{Q}}=i(P_{ABB'A'\mathcal{Q}}-P_{BAA'B'\mathcal{Q}})
\end{equation}
For reference, we also state the relations between corresponding
tensorial and spinorial objects of interest. The relation between
a 2-form $F_{ab}$ and a symmetric spinor $\varphi_{AB}$ is
$$
F_{ab}=\varphi_{AB}\bar\varepsilon_{A'B'}+\bar\varphi_{A'B'}\varepsilon_{AB}
\qquad ; \qquad \varphi_{AB}={1\over 2}F_{AC'B}{}^{C'}
$$
and one also has
$$
-F_{ac}F_{b}{}^c+{1\over 4}g_{ab}F_{cd}F^{cd}=2\varphi_{AB}\bar\varphi_{A'B'} \  .
$$
For the Weyl tensor $C_{abcd}$ and the completely symmetric Weyl spinor
$\Psi_{ABCD}$ the corresponding relations are
\begin{equation}\label{eq:w1}
C_{abcd}=\Psi_{ABCD}\bar\varepsilon_{A'B'}\bar\varepsilon_{C'D'}+
\bar\Psi_{A'B'C'D'}\varepsilon_{AB}\varepsilon_{CD}
\qquad ; \qquad \Psi_{ABCD}={1\over 4}C_{AE'B}{}^{E'}{}_{CF'D}{}^{F'}  
\end{equation}
and
\begin{equation}\label{eq:w2}
C_{akcl}C_b{}^k{}_d{}^l+{}^*C_{akcl}{}^*C_b{}^k{}_d{}^l=4\Psi_{ABCD}\bar\Psi_{A'B'C'D'} \ .
\end{equation}

\smallskip

That a tensor $T_{a\dots b}$ is completely symmetric and trace-free is very elegantly
 expressed in an equivalent way using spinor indices as
$$
T_{a\dots b}=T_{(A\dots B)(A'\dots B')} \ .
$$
We shall study when a tensor can be factorized in terms of spinors. If  a tensor $\tau_{a\dots b}$ 
 can be written
\begin{equation}\label{eq:fa}
\tau_{a\dots b}=\chi_{A\dots B}\bar\chi_{A'\dots B'} \ ,
\end{equation}
for some  spinor $\chi_{A\dots B}$, then it
follows  that $\tau_{a\dots b}$ satisfies the dominant property (\ref{dp}) and
 \begin{equation}\label{eq:fun}
\tau_{A\dots B}^{A'\dots B'}\tau_{C\dots D}^{C'\dots D'}=
 \tau_{A\dots B}^{C'\dots D'}\tau_{C\dots D}^{A'\dots B'} \ .
\end{equation}
 Conversely, suppose that $\tau_{a\dots b}$ satisfies (\ref{eq:fun}). Let $u^a,\dots , v^a$
 be future-directed null vectors such that $\tau_{a\dots b}u^a\dots v^b=k\ne 0$. Such
 null vectors must exist since otherwise, by taking linear combinations,  we would get
 $\tau_{a\dots b}u^a\dots v^b=0$ for all
 vectors which would imply $\tau_{a\dots b}=0$ . Then write the null vectors in terms
 of spinors as $u^a=\alpha^A\bar\alpha^{A'}, \dots ,v^a=\beta^A\bar\beta^{A'}$.
 Contract  (\ref{eq:fun}) with these spinors to get
 $$
 \tau_{A\dots BA'\dots B'}\tau_{C\dots DC'\dots D'}\alpha^C\bar\alpha^{C'}\dots
  \beta^D\bar\beta^{D'}=
 (\tau_{A\dots BC'\dots D'}\bar\alpha^{C'}\dots\bar\beta^{D'})(
 \tau_{C\dots DA'\dots B'}\alpha^{C}\dots\beta^{D})
 $$
 from which follows that  $\tau_{a\dots b}$ and $-\tau_{a\dots b}$ can be
 factorized as in (\ref{eq:fa}), one of them with $\chi_{A\dots B}={1\over \sqrt{|k|}}\tau_{A\dots BC'\dots
 D'}\bar\alpha^{C'}\dots\bar\beta^{D'}$ and the other with an extra $i$ in the factor,
 and that either $\tau_{a\dots b}$ or $-\tau_{a\dots b}$ has
the dominant property.

Finally, we introduce the following useful notation
$$
T\cdot T=T_{a\dots b}T^{a\dots b}
$$
 for any tensor $T_{a\dots b}$.

\section{Differential conditions for even
rank}\label{sec:Diffevenrank}

Suppose the tensor $T_{a_1...a_r}$, with $r$ even, can be
factorized according to

$$
    T_{a_1...a_r}=\Psi_{A_1...A_r}\bar{\Psi}_{A_1'...A_r'}
$$
with $\Psi_{A_1...A_r}$ symmetric. Then $T_{a_1...a_r}$ is
symmetric, trace-free and satisfies the dominant property. Note that
$T_{a_1...a_r}$ is invariant under $\Psi_{A_1...A_r}\to e^{-i\theta}\Psi_{A_1...A_r}$.
 We now prove a generalization of Theorem  \ref{ClassDiffRai} (or Theorem
 \ref{ClassDiffRaiSp}).

\begin{theorem}     \label{th:evenindexRainich}
Let $r$ be even and suppose that $\ \nabla^{a_1}T_{a_1...a_r}=0\neq T\cdot T\ $ and
$\ T_{a_1...a_r}= \Phi_{A_1...A_r}\bar\Phi_{A_1'...A_r'}$ for some
 totally symmetric $\Phi_{A_1...A_r}$. Then
$T_{a_1...a_r}= \Psi_{A_1...A_r}\bar\Psi_{A_1'...A_r'}$ for some
 totally symmetric $\Psi_{A_1...A_r}$ satisfying
$\nabla^{A_1A_1'}\Psi_{A_1...A_r}=0$  if and only if
$$\nabla_aS_b=\nabla_bS_a, \qquad {\mbox{where}}\qquad
S_b=\frac{e_{a_1b}{}^{pq}T_{p a_2...a_{r}}\nabla_q
T^{a_1a_2...a_r}}{T\cdot T}
$$
\end{theorem}
\begin{proof}
Since $T_{a_1...a_r}=\Phi_{A_1...A_r}\bar\Phi_{A_1'...A_r'}$ is preserved under "rotations"
$\Phi_{A_1...A_r} \to e^{i\chi}\Phi_{A_1...A_r}$ ($\chi$ real), we may assume that
$$
K=\frac{1}{2}\Phi_{A_1...A_r}\Phi^{A_1...A_r}
$$
is real (otherwise rotate $\Phi_{A_1...A_r}$ with a suitable $\chi$).

Now, we want to find the condition for the existence of some $\Psi_{A_1...A_r}$ with
$\Psi_{A_1...A_r}=\Psi_{(A_1...A_r)}$, $T_{a_1...a_r}=\Psi_{A_1...A_r}\bar\Psi_{A_1'...A_r'}$ and
$\nabla^{A_1A_1'}\Psi_{A_1...A_r}=0$. Clearly we can write $\Psi_{A_1...A_r}=e^{-i\theta}\Phi_{A_1...A_r}$
for some real $\theta$ with $\Phi_{A_1...A_r}$ as above.
If $\Psi_{A_1...A_r}$ satisfies the given field equations we have
(using the Leibniz rule)

$$
\nabla_{A_1A_1'}(e^{-i\theta}\Phi^{A_1...A_r})=e^{-i\theta}\nabla_{A_1A_1'}\Phi^{A_1...A_r}-ie^{-i\theta}\Phi^{A_1...A_r}\nabla_{A_1A_1'}\theta=0
$$
Cancelling the $e^{-i\theta}$ and contracting with
$\Phi_{BA_2...A_r}$ we get

$$
    \Phi_{BA_2...A_r}\nabla_{A_1A_1'}\Phi^{A_1A_2...A_r}-i\Phi_{BA_2...A_r}\Phi^{A_1A_2...A_r}\nabla_{A_1A_1'}\theta=0
$$
Using (\ref{eq:sas2}), (\ref{eq:zz}) and the fact that $r$ is even we have
 $$
 \Phi_{BA_2...A_r}\Phi^{A_1A_2...A_r}=\varepsilon_B{}^{A_1}K
 $$
 so we arrive at

$$
    \Phi_{BA_2...A_r}\nabla_{A_1A_1'}\Phi^{A_1 A_2...A_r}-iK\nabla_{BA_1'}\theta=0
$$
Relabeling $A_1$ and $B$ we get

$$
    \nabla_{A_1A_1'}\theta=\frac{1}{iK}\Phi_{A_1 A_2...A_r}\nabla_{BA_1'}\Phi^{BA_2...A_r}
$$
If we define a vector
\begin{equation}    \label{Condition4theta}
 S_{A_1 A_1'}=\frac{1}{iK}\Phi_{A_1 A_2...A_r}\nabla_{BA_1'}\Phi^{BA_2...A_r}
\end{equation}
then, expanding 
 $\ \nabla^{b}T_{ba_2...a_r}=\nabla^{BB'}(\Phi_{BA_2...A_r}\bar\Phi_{B' A_2'...A_r'})=0\ $ 
 by Leibniz' rule and contracting by  $\Phi^{A_1 A_2...A_r}\bar\Phi^{A_1' A_2'...A_r'}$ we get
 $$
 \bar\varepsilon_{B'}{}^{A_1'}K\Phi_{A_1 A_2...A_r}\nabla_{BB'}\Phi^{B A_2...A_r}+
 \varepsilon_{B}{}^{A_1}K\bar\Phi_{A_1' A_2'...A_r'}\nabla_{BB'}\bar\Phi^{B' A_2'...A_r'}=0
 $$
 or
  $$
 \Phi_{A_1 A_2...A_r}\nabla_{B A_1'}\Phi^{B A_2...A_r}+
\bar\Phi_{A_1' A_2'...A_r'}\nabla_{A_1 B'}\bar\Phi^{B' A_2'...A_r'}=0 \ ,
 $$ 
 hence the vector $\Phi_{A_1 A_2...A_r}\nabla_{B A_1'}\Phi^{B A_2...A_r}$ is purely
 imaginary and therefore $S_a$  is a real vector. 

We want to translate the right hand side of (\ref{Condition4theta}) into a tensorial expression.
Differentiate the tensor
$T_{a_1...a_r}=\Phi_{A_1...A_r}\bar{\Phi}_{A_1'...A_r'}$ and make
one contraction, leading to

$$
\nabla_{A_1B'}T^{A_1A_1'a_2...a_r}=\bar{\Phi}^{A_1'...A_r'}\nabla_{A_1B'}\Phi^{A_1...A_r}+\Phi^{A_1...A_r}\nabla_{A_1B'}\bar\Phi^{A_1'...A_r'}
$$
If we contract this with $T_{BA_1'A_2A_2'...A_rA_r'}$ we get, again using that $r$ is even,

$$
    T_{BA_1'a_2...a_r}\nabla_{A_1B'}T^{A_1A_1'a_2...a_r}=
2\Phi_{BA_2...A_r}K\nabla_{A_1B'}\Phi^{A_1A_2...A_r}+
\varepsilon_B{}^{A_1}K\nabla_{A_1B'}K
$$
Now we can use (\ref{Condition4theta}) to get

$$
    T_{BA_1'a_2...a_r}\nabla_{A_1B'}T^{A_1A_1'a_2...a_r}=2iK^2 S_{BB'}+K\nabla_{BB'}K
$$
On the right-hand side the first term is purely imaginary and the second is real, so taking the complex conjugate and then taking the difference results in

$$
    T_{BA_1'a_2...a_r}\nabla_{A_1B'}T^{A_1A_1'a_2...a_r}-T_{B'A_1a_2...a_r}\nabla_{BA_1'}T^{A_1A_1'a_2...a_r}=4iK^2 S_{BB'}
$$
Finally we can use (\ref{changepairofindices}) on the index pairs
$A_1A_1'$ and $BB'$ to get

$$
    4iK^2 S_{BB'} =ie_{a_1BB'}{}^{pq}T_{p a_2...a_r}\nabla_{q}T^{a_1a_2...a_r}
$$
Here $4K^2=T\cdot T$ so we get the  formula

\begin{equation}    \label{DiffS}
    S_b=\frac{e_{a_1b}{}^{pq}T_{p a_2...a_{r}}\nabla_qT^{a_1a_2...a_r}}{T\cdot T}
\end{equation}
Conversely, with $S_a$ given by (\ref{DiffS}), there is a real solution  
 $\theta$ (determined up to an additive constant) to the equation $\nabla_a\theta=S_a$  if 
  the integrability condition 
$$
\nabla_a S_b=\nabla_b S_a
$$
  is satisfied. This completes the proof.
 
\end{proof}

Note that  the above proof does not hold for odd $r$ in which
case $\Psi_{A_1...A_r}\Psi^{A_1...A_r}=0$ so $T\cdot T=0$
 as well.

\section{Complete Rainich theory for the Bel-Robinson tensor for Petrov types I, II and D}

As mentioned earlier, the algebraic Rainich condition for the
Bel-Robinson tensor was obtained in \cite{BL} but we restate the
result here

\begin{theorem}\label{th:RaiBR}
A completely symmetric and trace-free rank-4
tensor $T_{abcd}$  is, up to sign, a Bel-Robinson type tensor,
i.e. $\pm
T_{abcd}=C_{akcl}C_b{}^k{}_d{}^l+{}^*C_{akcl}{}^*C_b{}^k{}_d{}^l$
where $C_{abcd}$ has the same algebraic symmetries as the Weyl
tensor, if and only if
\begin{eqnarray}\label{eq:Rai}
T_{jabc}T^{jefg}=&{3\over 2}g_{(a}{}^{(e}T_{bc)jk}T^{fg)jk}+
{3\over 4}g_{(a}{}^{(e}T_{|jk|b}{}^{f}T_{c)}{}^{g)jk}
-{3\over 4}g_{(ab}T_{c)jk}{}^{(e}T^{fg)jk}
\nonumber \\
-&{3\over 4}g^{(ef}T_{jk(ab}T_{c)}{}^{g)jk}
+{1\over 32}(3g_{(ab}g_{c)}{}^{(e}g^{fg)}-4g_{(a}{}^{(e}g_b{}^f g_{c)}{}^{g)})T_{jklm}T^{jklm}
\end{eqnarray}
\end{theorem}
Equivalently this may also be stated as $T_{abcd}$ is the
superenergy tensor \cite{S} of a Weyl candidate tensor (that is a
tensor with same algebraic symmetries as the Weyl tensor:
$C_{abcd}=-C_{bacd}=-C_{abdc}=C_{cdab},\
C_{abcd}+C_{adbc}+C_{adcb}=0,\ C^a{}_{bad}=0$). 
As shown in \cite{BL} the identity (\ref{eq:Rai}) in Theorem \ref{th:RaiBR} can equivalently 
be replaced by
\begin{eqnarray*}\label{eq:RaiS}
T_{jbc(a}T_{e)}{}^{jfg}=&g_{(b}{}^{(f}T_{c)jk(a}T_{e)}{}^{g)jk}-
{1\over 4}g^{fg}T_{jkb(a}T_{e)c}{}^{jk}- {1\over
4}g_{bc}T_{jk}{}^{f}{}_{(a}T_{e)}{}^{gjk}
\nonumber \\
+&{1\over 4}g_{ae}(T_{jkbc}T^{jkfg}+{1\over
8}(g_{bc}g^{fg}-g_b{}^f g_c{}^g- g_b{}^g g_c{}^f
)T\cdot T)
\end{eqnarray*}

In terms of spinors we can state Theorem \ref{th:RaiBR} as

\begin{theorem}\label{th:RaiBR2}
A completely symmetric and trace-free rank-4 tensor $T_{abcd}$
can be written $\pm T_{abcd}=\Psi_{ABCD}\bar\Psi_{A'B'C'D'}$ with
$\Psi_{ABCD}=\Psi_{(ABCD)}$  if and only if (\ref{eq:Rai}) is satisfied.
\end{theorem}
Thus from a spinorial viewpoint this is a natural generalization
of the classical Rainich theory. We can then ask the same question
as in the classical case, e.g. what is required in order to have
$C_{abcd}$ (or $\Psi_{ABCD}$) satisfy some field equations? In
this case we choose the source-free gravitational field equations
\begin{equation}    \label{FieldEquations}
    \nabla^{AA'}\Psi_{ABCD}=0
\end{equation}
for the Weyl spinor that hold whenever Einstein's vacuum equations
hold. The tensor form of the equation (\ref{FieldEquations}) is
the vacuum Bianchi identity $\nabla_{[a}C_{bc]de}=0$ 
($\Leftrightarrow \nabla^a C_{abcd}=0$ in four dimensions)
for the Weyl tensor. 
From Theorem
\ref{th:evenindexRainich}, we immediately have the following
generalization of Theorem \ref{ClassDiffRai},

\begin{corollary}\label{cor:d4}
If $T_{abcd}=\Phi_{ABCD}\bar{\Phi}_{A'B'C'D'}$
for a completely symmetric spinor $\Phi_{ABCD}$ and if $\nabla^aT_{abcd}=0$, 
then in a region
where $T\cdot T\neq 0$ we have $T_{abcd}=\Psi_{ABCD}\bar{\Psi}_{A'B'C'D'}$
for a completely symmetric spinor $\Psi_{ABCD}$ satisfying
$\nabla^{AA'}\Psi_{ABCD}=0$ if and
only if 
\begin{equation}\label{di4}
\nabla_aS_b=\nabla_bS_a, \qquad {\mbox{where}}\qquad
S_e=\frac{e_{ae}{}^{pq}T_{bcdp}\nabla_q
T^{abcd}}{T\cdot T}
\end{equation}
\end{corollary}
This corollary gives a differential Rainich like
condition on the Bel-Robinson tensor. 
Combining Theorem \ref{th:RaiBR} (or \ref{th:RaiBR2}) and Corollary \ref{cor:d4}
we get the rank-4 generalization of Corollary \ref{ClassRai} which gives the
complete Rainich theory for Bel-Robinson type tensors. 
The tensor version is
\begin{corollary}\label{cor:Rai4}
Suppose that $T_{abcd}$ is completely symmetric, trace-free and divergence-free and that
$T\cdot T\neq 0$. Then $\pm T_{abcd}=C_{akcl}C_b{}^k{}_d{}^l+{}^*C_{akcl}{}^*C_b{}^k{}_d{}^l$
for a Weyl candidate tensor $C_{abcd}$ satisfying
$\nabla_{[a}C_{bc]de}=0$ if and only if (\ref{eq:Rai}) and (\ref{di4}) are satisfied.
\end{corollary}
Expressed in terms of spinors we get
\begin{corollary}\label{cor:Rai4s}
Suppose that $T_{abcd}$ is completely symmetric, trace-free and divergence-free and that
$T\cdot T\neq 0$. Then $\pm T_{abcd}=\Psi_{ABCD}\bar{\Psi}_{A'B'C'D'}$
for a completely symmetric spinor $\Psi_{ABCD}$ satisfying
$\nabla^{AA'}\Psi_{ABCD}=0$ if and only if (\ref{eq:Rai}) and (\ref{di4}) are satisfied.
\end{corollary}

We now proceed to see when these conditions imply that $C_{abcd}$ is not only
a Weyl {\it{candidate}} tensor satisfying
$\nabla_{[a}C_{bc]de}=0$ but the actual Weyl tensor of the spacetime.
 First of all, $T\cdot T=0$ if and only if
the spacetime is of Petrov type $III$ or $N$ \cite{B2}. Thus  we restrict ourselves
 to spacetimes of Petrov type $I,II$ and $D$. Bell and Szekeres
\cite{BellSzekeres} call a spacetime in which  (\ref{FieldEquations}) is satisfied
 by the actual Weyl spinor a {\it{C-space}}, hence all vacuum spacetimes are
 C-spaces.  This is also equivalent to the vanishing of the Cotton tensor
 $C_{abc}=2\nabla_{[a}R_{b]c}+{1\over 3}g_{c[a}\nabla_{b]}R\ $ \cite{GHHM}.  
 For spacetimes of Petrov type $I$, Bell and Szekeres prove 

\begin{theorem}
In an algebraically general C-space the source free field
equations (\ref{FieldEquations}) (the vacuum Bianchi identities) have a 
 unique solution to
within constant multiples, or its solutions are linear
combinations of at most two independent solutions.
\end{theorem}
 In \cite{BellSzekeres} conditions for the cases with non-unique
 solutions are given and the authors claim that most physically acceptable 
 metrics do not satisfy these conditions. As the conditions are not so simply stated,
 we refer to \cite{BellSzekeres} for further discussion. With the exception of these cases,
there is, up to a multiplicative constant, a unique solution to
(\ref{FieldEquations}) which then is of course the Weyl
spinor (so the gravitational field is uniquely determined by the Bianchi identities).  
 
 For Petrov types $II$ and $D$, let $o_A,\iota_A$ be a spin basis, 
 such that $o_A$ is the repeated principal
 null direction in spacetimes of Petrov type $II$, and such that $o_A$ and $\iota_A$ 
 are the repeated principal  null directions in spacetimes of Petrov type $D$,
and let, for the remaining of this section, $\Psi_{ABCD}$ denote the actual Weyl spinor
 of spacetime.
 Then  the following was proved in \cite{BellSzekeres}

\begin{theorem}
In a C-space of Petrov type $II$, the solution $\Phi_{ABCD}$ of
the source free field equations  (\ref{FieldEquations}) is unique up to a constant
$\alpha$ and null type fields $N_{ABCD}^1=\beta o_Ao_Bo_Co_D$ with
 $\beta$ a scalar,
according to $\Phi_{ABCD}=\alpha\Psi_{ABCD}+N_{ABCD}^1$ where 
 $\Psi_{ABCD}$ is the Weyl spinor.  For
Petrov type $D$ the solution can be written
$\Phi_{ABCD}=\alpha\Psi_{ABCD}+N_{ABCD}^1+N_{ABCD}^2$, where
$N^2_{ABCD}=\gamma\iota_A\iota_B\iota_C\iota_D$ with $\gamma$ a scalar.
\end{theorem}
 In deriving these theorems Bell and Szekeres use the Buchdahl conditions
 \cite{PR}
 $$
 \Psi^{ABC}{}_{(D}\Phi_{E\dots F)ABC}=0
 $$
 which are algebraic consistency conditions that relate any solution $\Phi_{A_1\dots A_n}$
 of the spin $n\over 2$-equation $\nabla^{A_1 A_1'}\Phi_{A_1\dots A_n}=0$ to
 the Weyl spinor $\Psi_{ABCD}$.
 Using these results we have

\begin{corollary}
In  C-spaces (including  vacuum spacetimes), 
  if $T_{abcd}$ is completely symmetric, trace-free and divergence-free,
 then, generically (Petrov type $I$ and excluding the exceptions given in  
 \cite{BellSzekeres}) and up to a constant factor, $T_{abcd}$ is  the Bel-Robinson 
 tensor of spacetime if and  only if (\ref{eq:Rai}) and (\ref{di4}) are satisfied.
\end{corollary}
For Petrov types $II$ and $D$ the following weaker conclusion can be drawn

\begin{corollary}
In   C-spaces (including  vacuum spacetimes), 
 if $T_{abcd}$ is completely symmetric, trace-free and divergence-free and if 
 spacetime is of Petrov type $II$ ($D$), then $T_{abcd}=\chi_{ABCD}\bar
 \chi_{A'B'C'D'}$ where 
 $\chi_{ABCD}=\alpha\Psi_{ABCD}+N_{ABCD}^1$ ($\chi_{ABCD}=\alpha\Psi_{ABCD}+N_{ABCD}^1+N_{ABCD}^2$)
 if and  only if (\ref{eq:Rai}) and (\ref{di4}) are satisfied.
\end{corollary}
 Note that the freedom in these cases does not preserve the principal null directions or even
the Petrov type.

\section{Algebraic conditions for rank 3}

In Senovilla's original definition of superenergy tensors of arbitrary tensors \cite{S},
all superenergy tensors are of even rank. However, in \cite{BS0} tensors of the form
$\Psi_{ABC}\bar\Psi_{A'B'C'}$ were used to study causal propagation of spin-${3\over 2}$
fields. In \cite{LP} Senovilla's definition has been extended to include superenergy
tensors of spinors and these may be of odd rank. Then, for instance, the superenergy tensor
of a completely symmetric spinor $\Psi_{A_1...A_r}$ of arbitrary rank is $T_{a_1...a_r}=\Psi_{A_1...A_r}\bar{\Psi}_{A_1'...A_r'}$. We now go on and study Rainich
type conditions for the rank-3 case, beginning with an algebraic characterization.

\begin{theorem}\label{th:alg3}
A completely symmetric and trace-free rank-3 tensor
$T_{abc}$  can be written $\pm T_{abc}=\Psi_{ABC}\bar\Psi_{A'B'C'}$
with $\Psi_{ABC}=\Psi_{(ABC)}$  if and only if
\begin{equation}\label{eq:alg3}
T_{abj}T^{dej}=g_{(a}{}^{(d}T_{b)jk}T^{e)jk}-{1\over 4}g_{ab}T^{d}{}_{jk}T^{ejk}-
{1\over 4}g^{de}T_{ajk}T_{b}{}^{jk}
\end{equation}
\end{theorem}

\begin{proof}
By the results in Section \ref{sec:spinors}  we must prove that (\ref{eq:alg3}) is
equivalent to
\begin{equation}\label{fund3}
T_{ABC}^{A'B'C'}T_{DEF}^{D'E'F'}-T_{ABC}^{D'E'F'}T_{DEF}^{A'B'C'}=0
\end{equation}
We follow the method developed in  \cite{BL} and divide up the left hand side in
symmetric and antisymmetric parts with respect
to the pairs $A'D'$, $B'E'$ and $C'F'$. Antisymmetric parts correspond to traces  so
for terms with 3, 2, 1 or 0 symmetrizations we have, respectively,
$$
\begin{array}{lll}
       &  T_{ABC}^{(C'|(B'|(A'}T_{DEF}^{D')|E')|F')}-
        T_{ABC}^{(F'|(E'|(D'}T_{DEF}^{A')|B')|C')}=0 \\
       & T_{J'ABC}^{(B'|(A'}T_{DEF}^{D')|E')J'}-
        T_{ABC}^{J'(E'|(D'}T_{DEFJ'}^{A')|B')}=
        2T_{J'ABC}^{(B'|(A'}T_{DEF}^{D')|E')J'} \\
        &   T_{J'K'ABC}^{(A'}T_{DEF}^{D')J'K'}-
        T_{ABC}^{J'K'(D'}T_{DEFJ'K'}^{A')}=0 \\
         & T_{J'K'L'ABC}T_{DEF}^{J'K'L'}-
        T_{ABC}^{J'K'L'}T_{DEFJ'K'L'}=
        2T_{J'K'L'ABC}T_{DEF}^{J'K'L'}
         \end{array}
$$
Therefore (\ref{fund3}) is equivalent to
$$
T_{J'ABC}^{(B'|(A'}T_{DEF}^{D')|E')J'}=0=T_{J'K'L'ABC}T_{DEF}^{J'K'L'}
$$
Now, continuing in the same way with respect to the unprimed indices of
these two expressions, expressions with an odd total number of contractions
vanish. Hence (\ref{fund3}) is equivalent to
\begin{equation}\label{dund31}
T_{j(B|(A}^{(B'|(A'}T_{D)|E)}^{D')|E')j}=0 \ , \
T_{jKL}^{(B'|(A'}T^{D')|E')jKL}=0 \ , \
T_{jK'L'(B|(A}T_{D)|E)}^{jK'L'}=0 \ , \
T_{jkl}T^{jkl}=0
\end{equation}
Now divide $T_{jBA}^{B'A'}T_{DE}^{D'E'j}$ up into symmetric and
antisymmetric parts four times in the index pairs $A'D'$, $AD$,
$BE$ and $B'E'$. Again, terms with an odd number of contractions
vanish and we get
$$
\begin{array}{lll}
       &  T_{jBA}^{B'A'}T_{DE}^{D'E'j}=
       T_{j(B|(A}^{(B'|(A'}T_{D)|E)}^{D')|E')j}+
      {1\over 4}\varepsilon_{BE}\varepsilon_{AD} T_{jKL}^{(B'|(A'}T^{D')|E')jKL}   \\
      & + {1\over 4}\bar\varepsilon^{B'E'}\bar\varepsilon^{A'D'}
      T_{jK'L'(B|(A}T_{D)|E)}^{jK'L'}
        + {1\over 4}\varepsilon_{BE}\bar\varepsilon^{B'E'}
       T_{jk(A}^{(A'}T_{D)}^{D')jk} \\
       & +{1\over 4}\varepsilon_{AD}\bar\varepsilon^{A'D'}
       T_{jk(B}^{(B'}T_{E)}^{E')jk}+
        {1\over 4}\varepsilon_{BE}\bar\varepsilon^{A'D'}
       T_{jk(A}^{(B'}T_{D)}^{E')jk} \\
       & +{1\over 4}\varepsilon_{AD}\bar\varepsilon^{B'E'}
       T_{jk(B}^{(A'}T_{E)}^{D')jk}+
        {1\over 16}\varepsilon_{BE}\varepsilon_{AD}\bar\varepsilon^{B'E'}
        \bar\varepsilon^{A'D'} T_{jkl}T^{jkl}
         \end{array}
$$
Since an expression is zero if and only if all its symmetric and antisymmetric parts are zero
we get that (\ref{fund3}) is equivalent to
\begin{equation}\label{fund32}
\begin{array}{lll}
         T_{jBA}^{B'A'}T_{DE}^{D'E'j}&=
       {1\over 4}\varepsilon_{BE}\bar\varepsilon^{B'E'}T_{jk(A}^{(A'}T_{D)}^{D')jk}
        +{1\over 4}\varepsilon_{AD}\bar\varepsilon^{A'D'}
       T_{jk(B}^{(B'}T_{E)}^{E')jk} \\
       & +        {1\over 4}\varepsilon_{BE}\bar\varepsilon^{A'D'}
       T_{jk(A}^{(B'}T_{D)}^{E')jk}
       +{1\over 4}\varepsilon_{AD}\bar\varepsilon^{B'E'}
       T_{jk(B}^{(A'}T_{E)}^{D')jk}
         \end{array}
\end{equation}
Now, note that, by using (\ref{eq:sas2}) on $A'B'$
$$
\begin{array}{lll}
      \varepsilon_{BE}\bar\varepsilon^{A'D'} T_{jk(A}^{(B'}T_{D)}^{E')jk}
     & =      \varepsilon_{BE}\bar\varepsilon^{B'D'} T_{jk(A}^{(A'}T_{D)}^{E')jk}+
      \bar\varepsilon^{A'B'}\varepsilon_{BE}\bar\varepsilon_{M'}{}^{D'}
       T_{jk(A}^{(M'}T_{D)}^{E')jk} \\
       & = \varepsilon_{BE}\bar\varepsilon^{B'D'} T_{jk(A}^{(A'}T_{D)}^{E')jk}+
      \bar\varepsilon^{A'B'}\varepsilon_{BE}T_{jk(A}^{(D'}T_{D)}^{E')jk}
 \end{array}
$$
Applying (\ref{eq:sas2}) with respect to $DE$ in the first term and $AE$ in the second
we have
$$
\begin{array}{lll}
      \varepsilon_{BE}\bar\varepsilon^{A'D'} T_{jk(A}^{(B'}T_{D)}^{E')jk}
     &=      \varepsilon_{BD}\bar\varepsilon^{B'D'} T_{jk(A}^{(A'}T_{E)}^{E')jk}+
     \varepsilon_{DE}\varepsilon_{B}{}^{M}\bar\varepsilon^{B'D'} T_{jk(A}^{(A'}T_{M)}^{E')jk} \\
       & +  \varepsilon_{BA}\bar\varepsilon^{A'B'} T_{jk(E}^{(D'}T_{D)}^{E')jk}+
     \varepsilon_{AE}\varepsilon_{B}{}^{M}\bar\varepsilon^{A'B'} T_{jk(M}^{(E'}T_{D)}^{D')jk} \\
     &  =    \varepsilon_{BD}\bar\varepsilon^{B'D'} T_{jk(A}^{(A'}T_{E)}^{E')jk}+
     \varepsilon_{DE}\bar\varepsilon^{B'D'} T_{jk(A}^{(A'}T_{B)}^{E')jk} \\
       & -  \varepsilon_{AB}\bar\varepsilon^{A'B'} T_{jk(D}^{(D'}T_{E)}^{E')jk}+
     \varepsilon_{AE}\bar\varepsilon^{A'B'} T_{jk(B}^{(D'}T_{D)}^{E')jk}
 \end{array}
$$
In the same way, first acting on $DE$ and then on $A'B'$ and $B'D'$, we find
$$
\begin{array}{lll}
      \varepsilon_{AD}\bar\varepsilon^{B'E'} T_{jk(B}^{(A'}T_{E)}^{D')jk}
     &  =    \varepsilon_{AE}\bar\varepsilon^{A'E'} T_{jk(B}^{(B'}T_{D)}^{D')jk}-
     \varepsilon_{AE}\bar\varepsilon^{A'B'} T_{jk(B}^{(D'}T_{D)}^{E')jk} \\
       & -  \varepsilon_{DE}\bar\varepsilon^{D'E'} T_{jk(A}^{(A'}T_{B)}^{B')jk}-
     \varepsilon_{DE}\bar\varepsilon^{B'D'} T_{jk(A}^{(A'}T_{B)}^{E')jk}
 \end{array}
$$
Substituting these expressions into (\ref{fund32}) gives
$$
\begin{array}{lll}
         T_{jBA}^{B'A'}T_{DE}^{D'E'j}&=
       {1\over 4}\varepsilon_{BE}\bar\varepsilon^{B'E'}T_{jk(A}^{(A'}T_{D)}^{D')jk}
        +{1\over 4}\varepsilon_{AD}\bar\varepsilon^{A'D'}
       T_{jk(B}^{(B'}T_{E)}^{E')jk} +
        {1\over 4}\varepsilon_{AE}\bar\varepsilon^{A'E'} T_{jk(B}^{(B'}T_{D)}^{D')jk} \\
       & +  {1\over 4}   \varepsilon_{BD}\bar\varepsilon^{B'D'} T_{jk(A}^{(A'}T_{E)}^{E')jk}-
        {1\over 4}\varepsilon_{AB}\bar\varepsilon^{A'B'} T_{jk(D}^{(D'}T_{E)}^{E')jk}-
        {1\over 4}\varepsilon_{DE}\bar\varepsilon^{D'E'} T_{jk(A}^{(A'}T_{B)}^{B')jk}
         \end{array}
$$
Lowering indices, we use  (\ref{eq:ds}) to rewrite this to
\begin{equation}
\begin{array}{lll}\label{fund33}
         T_{jab}T_{de}{}^{j}=&
       {1\over 4}g_{be}T_{jka}T_{d}^{jk}  +{1\over 4}g_{ad}T_{jkb}T_{e}^{jk}  +
      {1\over 4}g_{ae}T_{jkb}T_{d}^{jk}   \\
       & +{1\over 4}g_{bd}T_{jka}T_{e}^{jk}-{1\over 4}g_{ab}T_{jkd}T_{e}^{jk}
        -{1\over 4}g_{de}T_{jka}T_{b}^{jk}
          \end{array}
\end{equation}
where we also used $T_{jkl}T^{jkl}=0$. Since (\ref{fund33}) is equivalent to
(\ref{fund3}) the proof is completed.

\end{proof}

\section{Differential conditions for rank 3}

It is clear that the methods of Section \ref{sec:Diffevenrank} do
not work for odd rank. We have e.g. that $T_{a\dots b}T^{a\dots
b}=0$ in this case and it is important if (\ref{eq:zz}) is used an
even or odd number of times. We present here a condition for rank
3 but it can be generalized to higher odd rank.
Given a completely symmetric spinor $\Psi_{ABC}$ we define a symmetric spinor
$$
\psi_{AB}=\Psi_{ACD}\Psi_B{}^{CD}
$$
Writing
$$
\Psi_{ABC}=\alpha_{(A}\beta_B\gamma_{C)}
$$
where $\alpha_{A}$, $\beta_{A}$ and $\gamma_{A}$ are the three principal
null directions of $\Psi_{ABC}$, we may say that $\Psi_{ABC}$ is of type I, II or
N if the principal null directions are all distinct, if two coincide, or if all three
coincide, respectively. It is then easy to see that $\psi_{AB}=0$ if and only if
$\Psi_{ABC}$ is of type N and that $\psi_{AB}\psi^{AB}\ne 0$ if and only if
$\Psi_{ABC}$ is of type I. With  $T_{abc}=\Psi_{ABC}\bar\Psi_{A'B'C'}$ we
see $T_{acd}T_b{}^{cd}T^a{}_{ef}T^{bef}\ne 0$ if and only if $\Psi_{ABC}$ is of
type I. For type I, the generic case, we have the following

\begin{theorem}\label{th:diff3}
Suppose that $T_{abc}=\Phi_{ABC}\bar\Phi_{A'B'C'}$ for some symmetric spinor
$\Phi_{ABC}$ and that $\nabla^aT_{abc}=0\ne T_{acd}T_b{}^{cd}T^a{}_{ef}T^{bef}$. 
 Then $T_{abc}=\Psi_{ABC}\bar\Psi_{A'B'C'}$ for some symmetric spinor
$\Psi_{ABC}$ satisfying  $\nabla^{AA'}\Psi_{ABC}=0$ if and only if
\begin{equation}\label{eq:diff3}
\nabla_a S_b =\nabla_b S_a \qquad {\mbox{where}}\qquad
S^h={{e^{haef}T_{emn}T^{bmn}T_b{}^{cd}\nabla_f T_{acd}}\over
{T_{acd}T_b{}^{cd}T^a{}_{ef}T^{bef}}}
\end{equation}
\end{theorem}

\begin{proof}

Since $T_{abc}=\Phi_{ABC}\bar\Phi_{A'B'C'}$ is preserved under "rotations"
$\Phi_{ABC} \to e^{i\chi}\Phi_{ABC}$ ($\chi$ real), we may assume that the symmetric
spinor $\phi_{AB}=\Phi_{ACD}\Phi_B{}^{CD}$ has the property that
$$
k=\phi_{AB}\phi^{AB}
$$
is real (otherwise rotate with a suitable $\chi$).
Now we want to find the condition for the existence of some $\Psi_{ABC}$ with
$\Psi_{ABC}=\Psi_{(ABC)}$, $T_{abc}=\Psi_{ABC}\bar\Psi_{A'B'C'}$ and
$\nabla^{AA'}\Psi_{ABC}=0$. Clearly we can write $\Psi_{ABC}=e^{-i\theta}\Phi_{ABC}$
for some real $\theta$. The differential equation becomes
$$
\nabla^{AA'}\Psi_{ABC}=\nabla^{AA'}(e^{-i\theta}\Phi_{ABC})=
e^{-i\theta}(\nabla^{AA'}\Phi_{ABC}-i\Phi_{ABC}\nabla^{AA'}\theta)=0
$$
Multiplying by $\Phi^D{}_{BC}$ we have
$$
\Phi^D{}_{BC}\nabla_{AA'}\Phi^{ABC}-i\phi^{AD}\nabla_{AA'}\theta=0
$$
Then multiply by $\phi_{DE}$ and use that $\phi_{D}{}^{E}\phi^{AD}$ is antisymmetric
in $AE$. This implies
$$
\phi_{DE}\Phi^D{}_{BC}\nabla_{AA'}\Phi^{ABC}=-{{i}\over 2}k\varepsilon_E{}^{A}
\nabla_{AA'}\theta=-{{i}\over 2}k\nabla_{EA'}\theta
$$
Hence
$$
\nabla_e\theta={{2i}\over k}\phi_{DE}\Phi^D{}_{BC}\nabla_{AE'}\Phi^{ABC}
$$
 Define a vector
\begin{equation}\label{d31}
 S_e={{2i}\over k}\phi_{DE}\Phi^D{}_{BC}\nabla_{AE'}\Phi^{ABC}
\end{equation}
 which is real since applying Leibniz' rule to $\nabla^aT_{abc}=
 \nabla^{AA'}(\Phi_{ABC}\bar\Phi_{A'B'C'})=0$, contracting with
$\phi_{DE}\bar\phi_{D'E'}\Phi^{DBC}\bar\Phi^{D'B'C'}$ and using
$2\phi_{AB}\phi^A{}_C=k\varepsilon_{BC}$, one finds that
the vector $\phi_{DE}\Phi^D{}_{BC}\nabla_{AE'}\Phi^{ABC}$ is
purely imaginary.

Next, translate the right hand side of (\ref{d31}) into a tensorial expression.
We have
$$
\begin{array}{lll}
 &T^{bmn}T_b{}^{cd}T^{HA'}{}_{mn}\nabla^{AH'}T_{acd} \\
&=T^{bmn}T_b{}^{cd}T^{HA'}{}_{mn}\nabla^{AH'}(\Phi_{ACD}\bar\Phi_{A'C'D'}) \\
&= \Phi^{BMN}\bar\Phi^{B'M'N'}\Phi_B{}^{CD}\bar\Phi_{B'}{}^{C'D'}
\Phi^H{}_{MN}\bar\Phi^{A'}{}_{M'N'}(\bar\Phi_{A'C'D'}\nabla^{AH'}\Phi_{ACD}+
\Phi_{ACD}\nabla^{AH'}\bar\Phi_{A'C'D'}) \\
&= \bar\phi^{A'B'}\bar\phi_{A'B'}\phi^{BH}\Phi_B{}^{CD}\nabla^{AH'}\Phi_{ACD}+
\phi^{BH}\phi_{AB}\bar\phi^{A'B'}\bar\Phi_{B'}{}^{C'D'}\nabla^{AH'}\bar\Phi_{A'C'D'} \\
 &=k(-{{ik}\over 2})S^{HH'}+{1\over 2}k\varepsilon_A{}^H
\bar\phi^{A'B'}\bar\Phi_{(B'}{}^{C'D'}\nabla^{AH'}\bar\Phi_{A')C'D'} \\
 &=-{{i}\over 2}k^2 S^{h}+{1\over 2}k\bar\phi^{A'B'}{1\over 2}
\nabla^{HH'}(\bar\Phi_{B'}{}^{C'D'}\bar\Phi_{A'C'D'}) \\
 &= -{{i}\over 2}k^2 S^{h}+{1\over 4}k\bar\phi^{A'B'}\nabla^{h}\bar\phi_{A'B'} \\
&= -{{i}\over 2}k^2 S^{h}+{1\over 8}k\nabla^{h}(\bar\phi_{A'B'}\bar\phi^{A'B'})\\
&= -{{i}\over 2}k^2 S^{h}+{1\over 8}k\nabla^{h}k
\end{array}
$$
Subtract the complex conjugate to get
$$
T^{bmn}T_b{}^{cd}(T^{HA'}{}_{mn}\nabla^{AH'}-T^{AH'}{}_{mn}\nabla^{HA'})T_{acd}=
-ik^2 S^{h}
$$
and apply (\ref{changepairofindices}) to get
\begin{equation}\label{D3S}
 S^{h}=-{1\over {k^2}}e^{ahef}T^{bmn}T_b{}^{cd}T_{emn}\nabla_f T_{acd}
\end{equation}
 Conversely, with the real vector $S_a$ given by (\ref{D3S}), the equation
 $\nabla_a\theta=S_a$ has a real solution $\theta$ (determined up to an
 additive constant) if the integrability condition 
$$
\nabla_a S_b=\nabla_b S_a 
$$
 is satisfied. This proves the theorem.

\end{proof}

\section{Complete Rainich theory for rank 3}

A symmetric rank-3 spinor $\Psi_{ABC}$ can be seen as representing
a spin-$\frac{3}{2}$ field on spacetime. The field equations for a
massless spin-$\frac{3}{2}$ field are

$$
    \nabla^{AA'}\Psi_{ABC}=0
$$
which are of the form in theorem \ref{th:diff3} above. Thus
collecting together the algebraic and differential conditions for
symmetric trace-free and divergence-free rank-3 tensors obtained above, we find the
following

\begin{theorem}     \label{th:CRai3}
Suppose that $T_{abc}$ is symmetric, trace-free, divergence-free and
that $T_{acd}T_b{}^{cd}T^a{}_{ef}T^{bef} \ne 0$. Then
$T_{abc}$ is the superenergy tensor of a massless
spin-$\frac{3}{2}$ field, i.e., $T_{abc}=\Psi_{ABC}\bar\Psi_{A'B'C'}$
for some symmetric spinor
$\Psi_{ABC}$ satisfying  $\nabla^{AA'}\Psi_{ABC}=0$, if and only if
$$
T_{abj}T^{dej}=g_{(a}{}^{(d}T_{b)jk}T^{e)jk}-{1\over
4}g_{ab}T^{d}{}_{jk}T^{ejk}- {1\over 4}g^{de}T_{ajk}T_{b}{}^{jk}
$$
and
$$
\nabla_a S_b =\nabla_b S_a \qquad {\mbox{where}}\qquad
S^h={{e^{haef}T_{emn}T^{bmn}T_b{}^{cd}\nabla_f T_{acd}}\over
{T_{acd}T_b{}^{cd}T^a{}_{ef}T^{bef}}}
$$
\end{theorem}

In analogy with the rank-2 and rank-4 cases, this can be seen as
a complete Rainich theory, in the mathematical sense since $T_{abc}$ is not
 linked directly to the geometry via the field equations in present physical theories, 
 for rank-3 superenergy tensors for  the generic (type I) case.

\section{Discussion}

We have presented a complete Rainich theory for superenergy
tensors of rank 3 and 4 in four dimensions in a generic case. However, the results
obtained may be generalized to higher rank superenergy tensors.
The interpretation is clear as the equations involved are the
equations for a massless spin-$\frac{n}{2}$ field. It is also
possible to pursue other generalizations of these results. For
example one could consider massive spin-$\frac{n}{2}$ fields, in
which case it is obviously necessary to modify the Theorems
\ref{th:evenindexRainich} and \ref{th:diff3}. One could also consider the
rank-4 differential conditions in spacetimes of Petrov type
$III$ and $N$, where $T\cdot T=0$ and Theorem
\ref{th:evenindexRainich} does not apply. 
From the results for the rank-2 case \cite{G,Lu} it is likely that this
case will be rather complicated and that it is not so easy to
apply Bell-Szekeres \cite{BellSzekeres} types of  results here
(which are already complicated for any algebraically special case).
Note, however, that the
algebraic conditions also apply to cases when $T\cdot T=0$. 
For generalizations to arbitrary spacetime dimension or to metrics of arbitrary 
signature tensor methods would be needed and it is clear that these
would be much more complicated than the spinor methods we have used
here.

\subsection*{Acknowledgements}
We thank Brian Edgar and Jos\'e Senovilla for useful suggestions and comments.

\end{document}